\documentclass[3p,times,procedia]{elsarticle}
\usepackage{nupha_ecrc}
\RequirePackage{lineno}

\volume{00}

\firstpage{1}

\journalname{Nuclear Physics A}

\runauth{V. Khachatryan for the PHENIX Collaboration}

\jid{nupha}

\jnltitlelogo{Nuclear Physics A}

\def\beq{\begin{equation}}
\def\eeq{\end{equation}}
\def\bea{\begin{eqnarray}}
\def\eea{\end{eqnarray}}

\newcommand{\Lb}{\left(}
\newcommand{\Rb}{\right)}
\newcommand \sqn{$\sqrt{s_{_{NN}}}$ } 

\usepackage[rightcaption]{sidecap}
\usepackage{graphicx,color,amsmath,amssymb}

\usepackage{amssymb}

\usepackage[figuresright]{rotating}

\begin{document}

\begin{frontmatter}



\dochead{XXVIIth International Conference on Ultrarelativistic Nucleus-Nucleus Collisions\\ (Quark Matter 2018)}

\title{PHENIX measurements of low momentum \\
direct photon radiation}


\author{Vladimir Khachatryan for the PHENIX Collaboration}

\address{Department of Physics and Astronomy, Stony Brook University, Stony Brook, New York 11794-3800, USA}

\begin{abstract}
The versatility of RHIC allowed the PHENIX collaboration to measure low momentum direct photons from small systems, such as p+p, p+A, d+Au at $\sqrt{s_{NN}} = $200\,GeV as well as from large A+A systems, such as Au+Au and Cu+Cu at 200\,GeV and Au+Au at 62.4\,GeV and 39\,GeV. In these measurements PHENIX has discovered a large excess over the scaled p+p yield of direct photons in A+A collisions, and a non-zero excess over the scaled p+p yield in central p+A collisions. Another PHENIX discovery is that at low-$p_{T}$ the integrated yield of direct photons, $dN_{\gamma}/dy$, from large systems follows a universal scaling as a function of the charged-particle multiplicity, $(dN_{ch}/d\eta)^{\alpha}$, with $\alpha = 1.25$. The observed scaling properties of direct photons from these systems show that the photon production yield increases faster than the charged-particle multiplicity.
\end{abstract}


\end{frontmatter}


\setlength{\linenumbersep}{6pt}

\section{Introduction}
\label{intro}
Direct photons are an important tool with unique capabilities to study the strongly interacting medium produced in (ultra)relativistic heavy ion collisions. By measuring these photons, one can gain information on the properties and dynamics of the produced matter integrated over space and time. By definition, the direct photons are the ``remnants" of subtraction of a large number of hadronic decay photons (mostly from $\pi^{0}$ and $\eta$ decays) from the total observed yield. They originate from the hot fireball of the Quark-Gluon Plasma (QGP), late hadronic phase as well as from initial hard scattering processes like QCD Compton scattering among the incoming and outgoing partons.
 
A quite challenging problem, dubbed as ``thermal photon puzzle", emerged when PHENIX measured large invariant yield and large anisotropy (elliptic flow) of low momentum direct photons in Au+Au collisions at \sqn = 200\,GeV \cite{Bannier:2016}. Various theoretical models encounter difficulties when they are used to describe these two quantities simultaneously though there is also some progress \cite{vanHees:2014ida,Paquet:2015lta,Kim:2016ylr}. In order to resolve this puzzle, PHENIX has measured low momentum direct photons in large and small collision systems. These measurements revealed very interesting findings reported in this article.

\section{Large and small systems: recent results on low momentum direct photon $p_{T}$ spectra}
\label{large}
\begin{figure}[h!]
\vspace{-3.5mm}
\begin{center}
\hspace{0.0\textwidth}
\vspace{0.0\textwidth}
   {\includegraphics[width=0.24\textwidth, height=0.23\textheight]{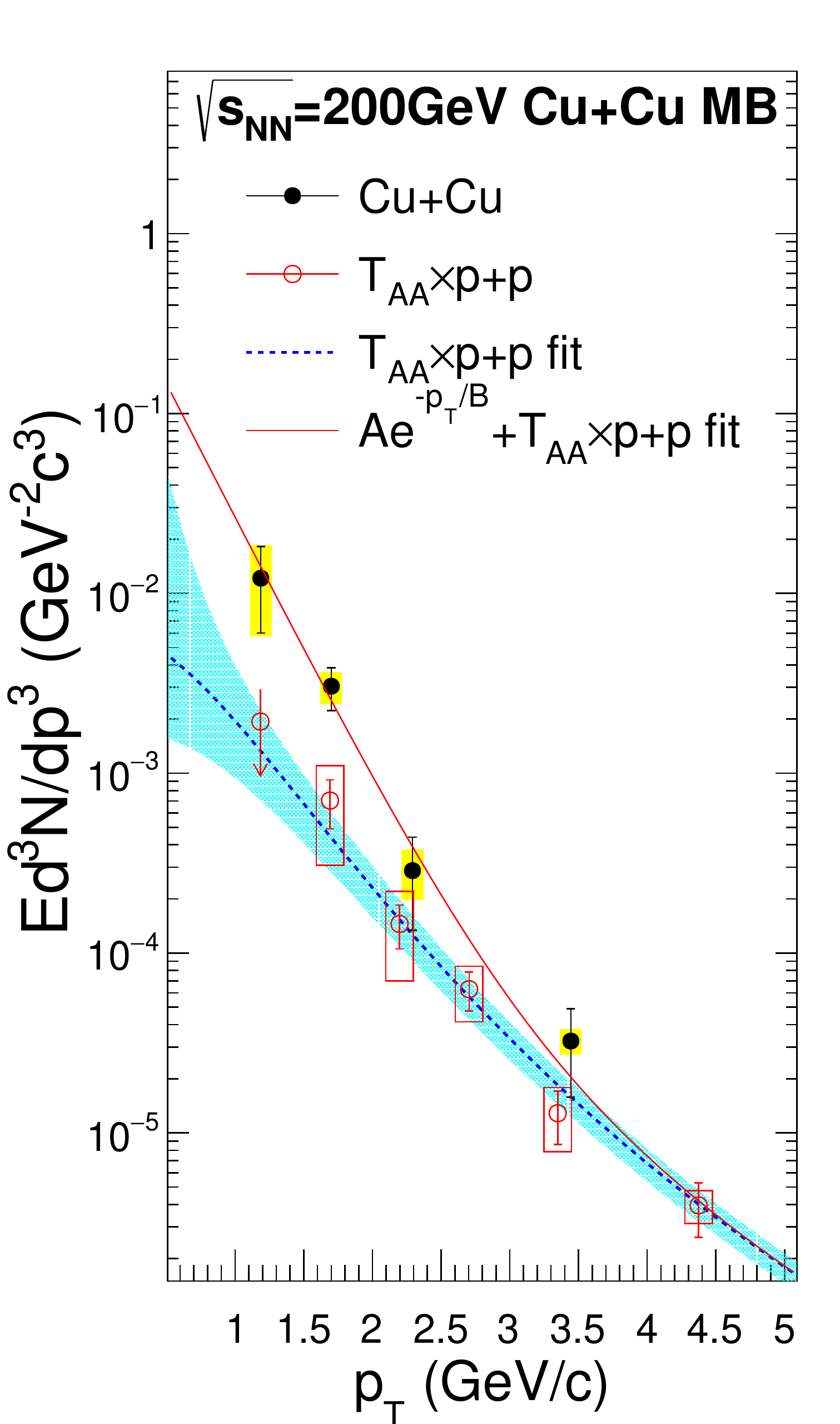} \label{fig:Fig1a}}
\hspace{-0.0175\textwidth}
\vspace{0.0\textwidth}
   {\includegraphics[width=0.375\textwidth]{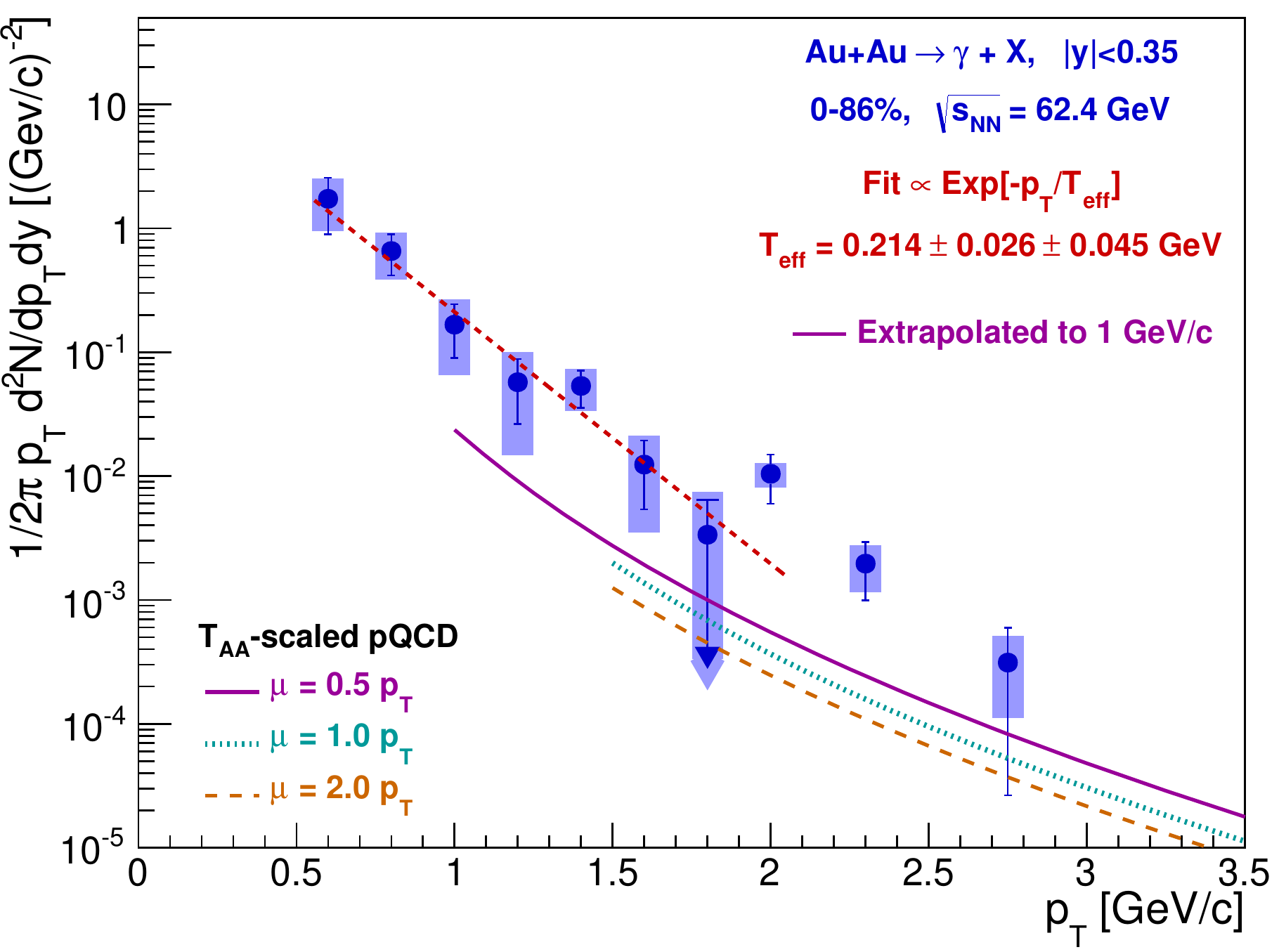} \label{fig:Fig1b}}
\hspace{-0.0175\textwidth}
\vspace{0.0\textwidth}
   {\includegraphics[width=0.375\textwidth]{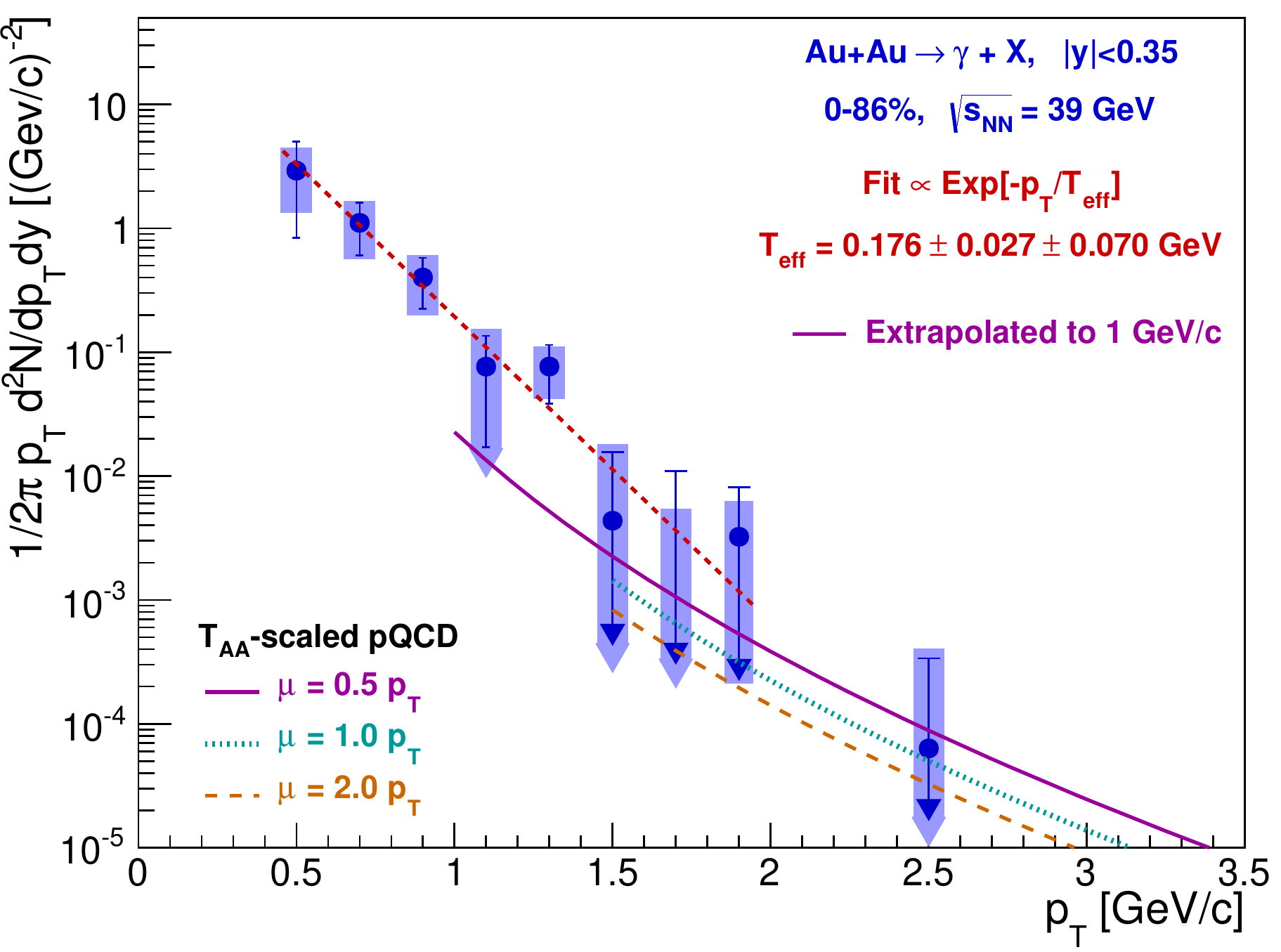} \label{fig:Fig1c}}
\end{center}
\vspace{-5.0mm}
\caption{Direct photon spectra at low-$p_{T}$ in Cu+Cu at \sqn = 200\,GeV \cite{Adare:cucu} (internal conversions) and in Au+Au at \sqn = 62.4\,GeV and 39\,GeV (external conversions with HBD) \cite{Adare:2018}. All data are in minimum bias.}
\label{fig:fig_large}
\end{figure}

Recently PHENIX accomplished low momentum direct photon measurements for large systems: Cu+Cu at \sqn = 200\,GeV through their internal conversions, and Au+Au at \sqn = 62.4\,GeV and 39\,GeV with the external conversion method. In Fig.\,\ref{fig:fig_large} one can see the results for minimum bias data samples. In the external conversion method the photons are measured through their conversions to electron-positron pairs at the HBD (or VTX) in the PHENIX detector system, and the fraction of direct photons is determined after tagging photons from neutral pion decays. Comparing these data to $T_{AA}$ scaled p+p fit or pQCD calculations one finds a significant excess over the scaled p+p yield of low $p_{T}$ direct photons in all three systems. 

\begin{figure}[h!]
\vspace{-3.5mm}
\begin{center}
\hspace{0.0\textwidth}
\vspace{0.0\textwidth}
   {\includegraphics[width=0.285\textwidth]{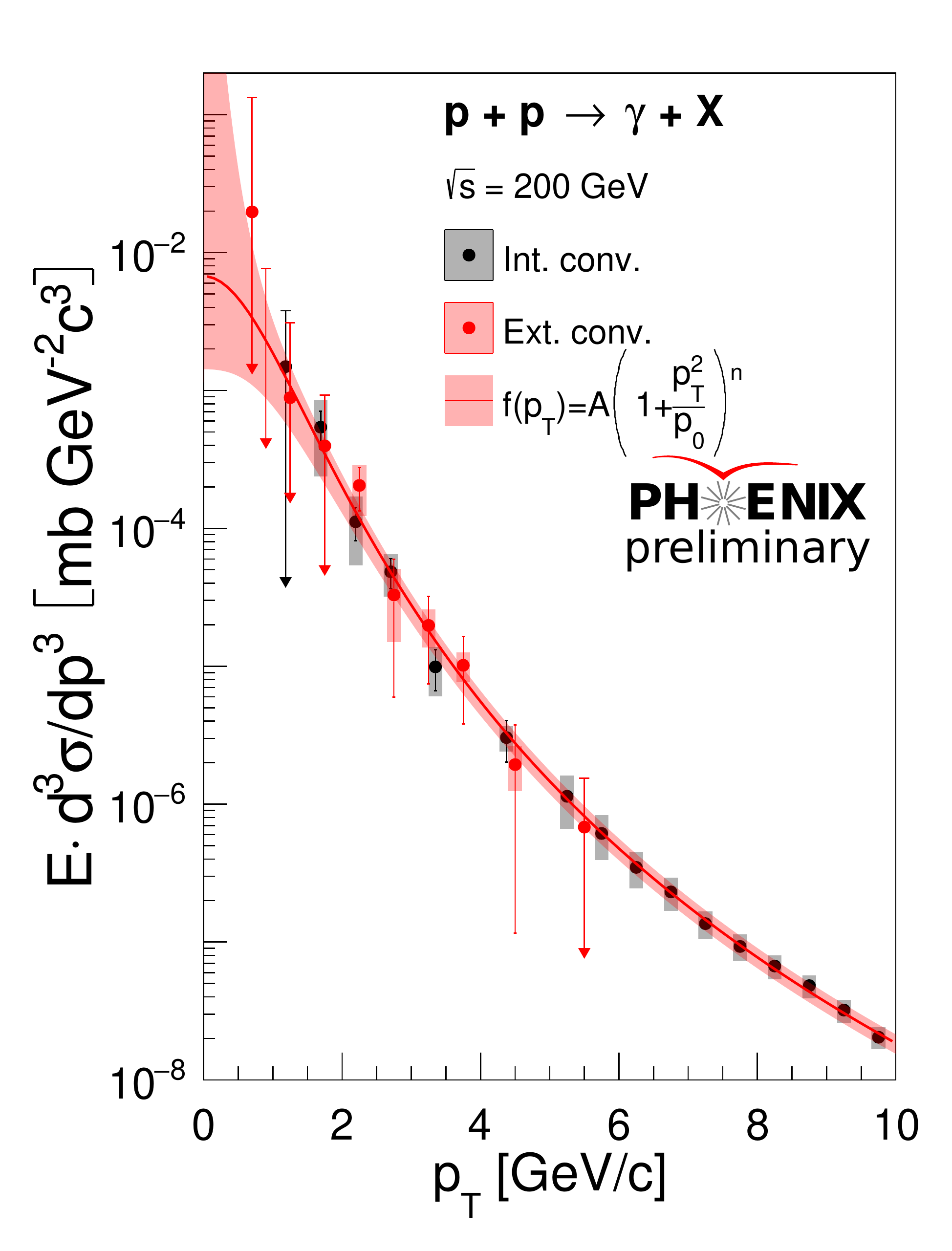} \label{fig:Fig2a}}
\hspace{0.0\textwidth}
\vspace{0.0\textwidth}
   {\includegraphics[width=0.285\textwidth]{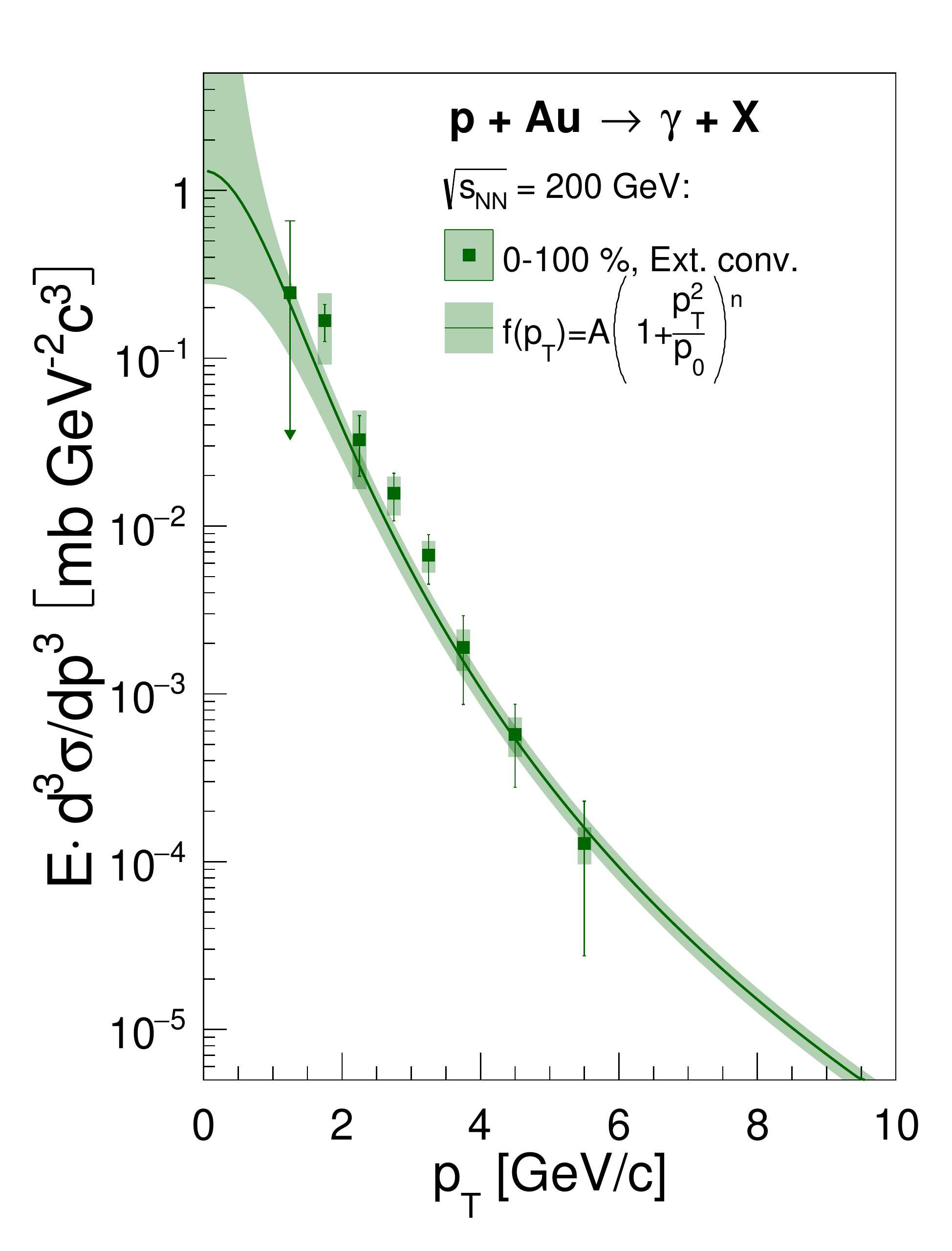} \label{fig:Fig2b}}
\hspace{0.0\textwidth}
\vspace{0.0\textwidth}
   {\includegraphics[width=0.285\textwidth]{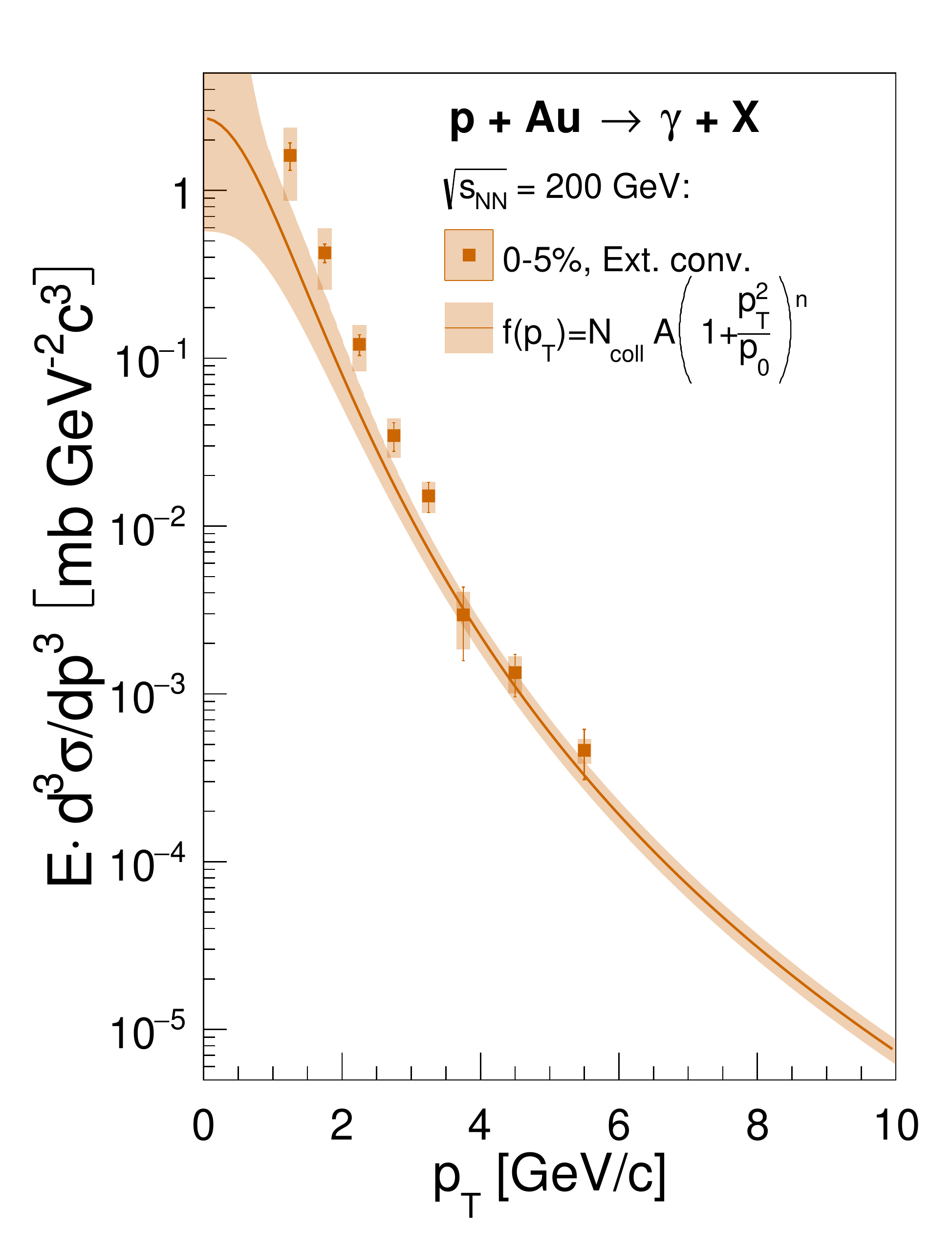} \label{fig:Fig2c}}
\end{center}
\vspace{-7.5mm}
\caption{Direct photon spectra at low-$p_{T}$ in p+p and in minimum bias and central p+Au collisions at 200\,GeV (external conversions with VTX). In the left plot the comparison with the PHENIX p+p results from internal conversions \cite{Adare:2013} is also shown.}
\label{fig:fig_small}
\end{figure}

With the external conversion method PHENIX recently also measured low momentum direct photons in p+p and p+A collisions (shown in Fig.\,\ref{fig:fig_small}). Within systematic uncertainties, the observed a non-zero excess yield ($\sim$ one sigma) in central p+Au collisions above the scaled p+p fit may come from the possible production of QGP droplets in small central systems.

\vspace{-0.02\textwidth}
\section{Direct photons scaling}
\label{scaling}
\begin{SCfigure}[][h!]
\vspace{0.0mm}
   \includegraphics[width=0.425\textwidth]{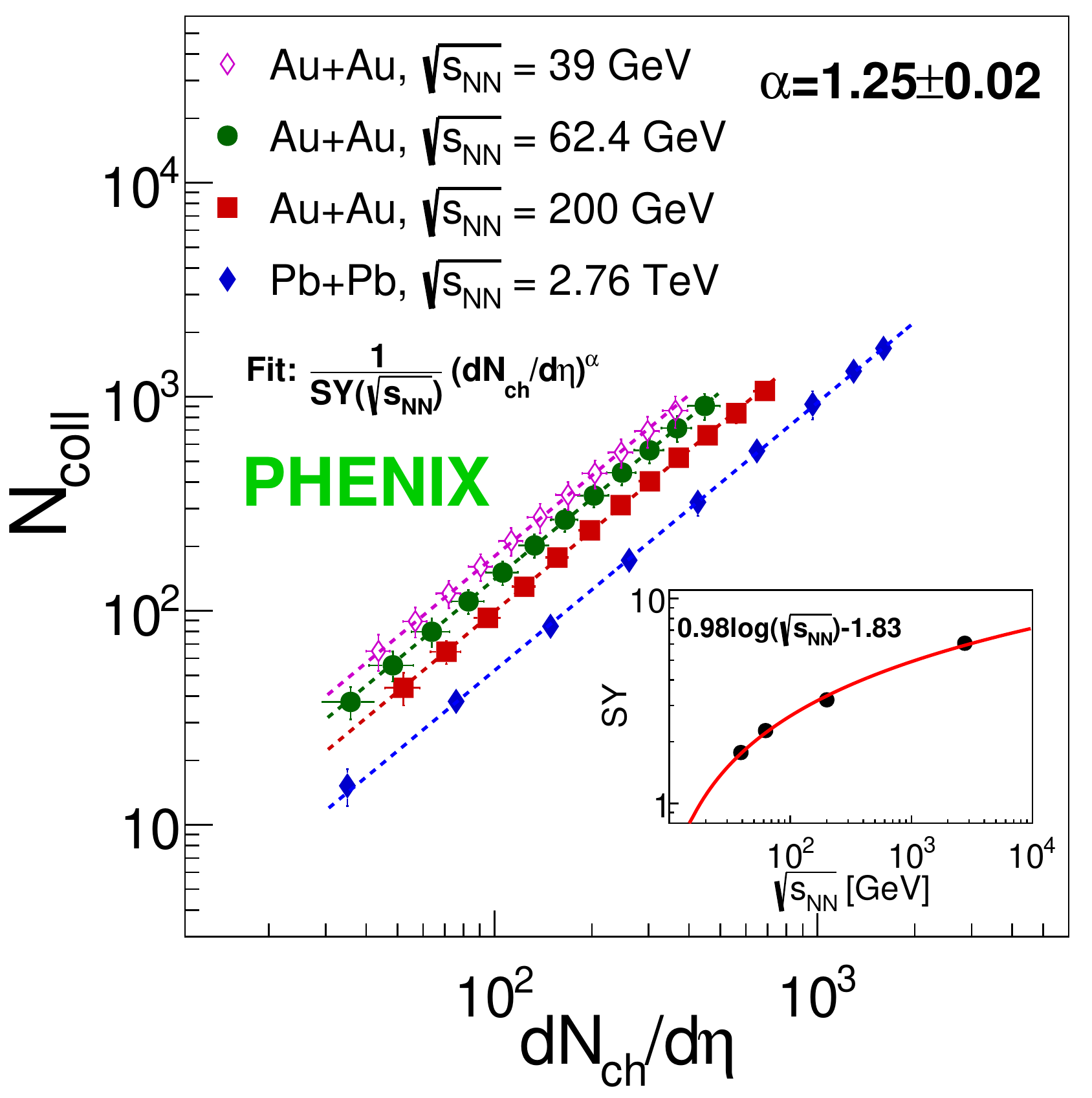}
\hspace{0.0mm}
\vspace{3.0mm}
\caption{The number of binary collisions, $N_{coll}$, vs. charged-particle  multiplicity, $dN_{ch}/d\eta|_{\eta=0}$, for four beam energies shown in the plot. All the data describing the large systems are simultaneously fitted by a power-law with $\alpha$ = 1.25, with horizontal and vertical error bars showing the uncertainties of $dN_{ch}/d\eta|_{\eta=0}$ \cite{Adare:nch,Aamodt:2011nch} and of $N_{coll}$ (from the Glauber Monte-Carlo simulations), respectively.

The small box on the bottom right shows data demonstrating a remarkable scaling between $N_{coll}$ and $dN_{ch}/d\eta$, which takes the form:
$
N_{coll} = \frac{1}{SY\!(\sqrt{s_{_{NN}}})} \Lb \frac{dN_{ch}}{d\eta} \Rb^{\alpha},
$
where the specific yield, SY, is introduced, which logarithmically increases with \sqn:
$
\mbox{SY}\!(\sqrt{s_{_{NN}}}) = 0.98\,\log(\sqrt{s_{_{NN}}}) - 1.83.
$
{\color{white} ......................................................................................................................................
........................................................................................................................................................
........................................................................................................................................................
........................................................................................................................................................
........................................................................................................................................................
}}
\label{fig:fig_scaling1}
\end{SCfigure}

For a given beam energy one can compare data from different centrality classes (or system size) using number of participants, $N_{part}$, or the number of binary collisions, $N_{coll}$. However, this is not useful to compare data at different energies. We therefore use charged-particle  multiplicity, $dN_{ch}/d\eta$, which itself has an interesting scaling behavior with $N_{coll}$ shown in Fig.\,\ref{fig:fig_scaling1}. Here $N_{coll}$ scales like $(dN_{ch}/d\eta)^{\alpha}$ for all \sqn with a logarithmically slowly increasing function called specific yield. The exponent $\alpha$ is found to be $\alpha = 1.25$. The other details are given in the caption of Fig.\,\ref{fig:fig_scaling1}.

\begin{figure}[h!]
\vspace{0.0mm}
\begin{center}
   {\includegraphics[width=0.85\textwidth]{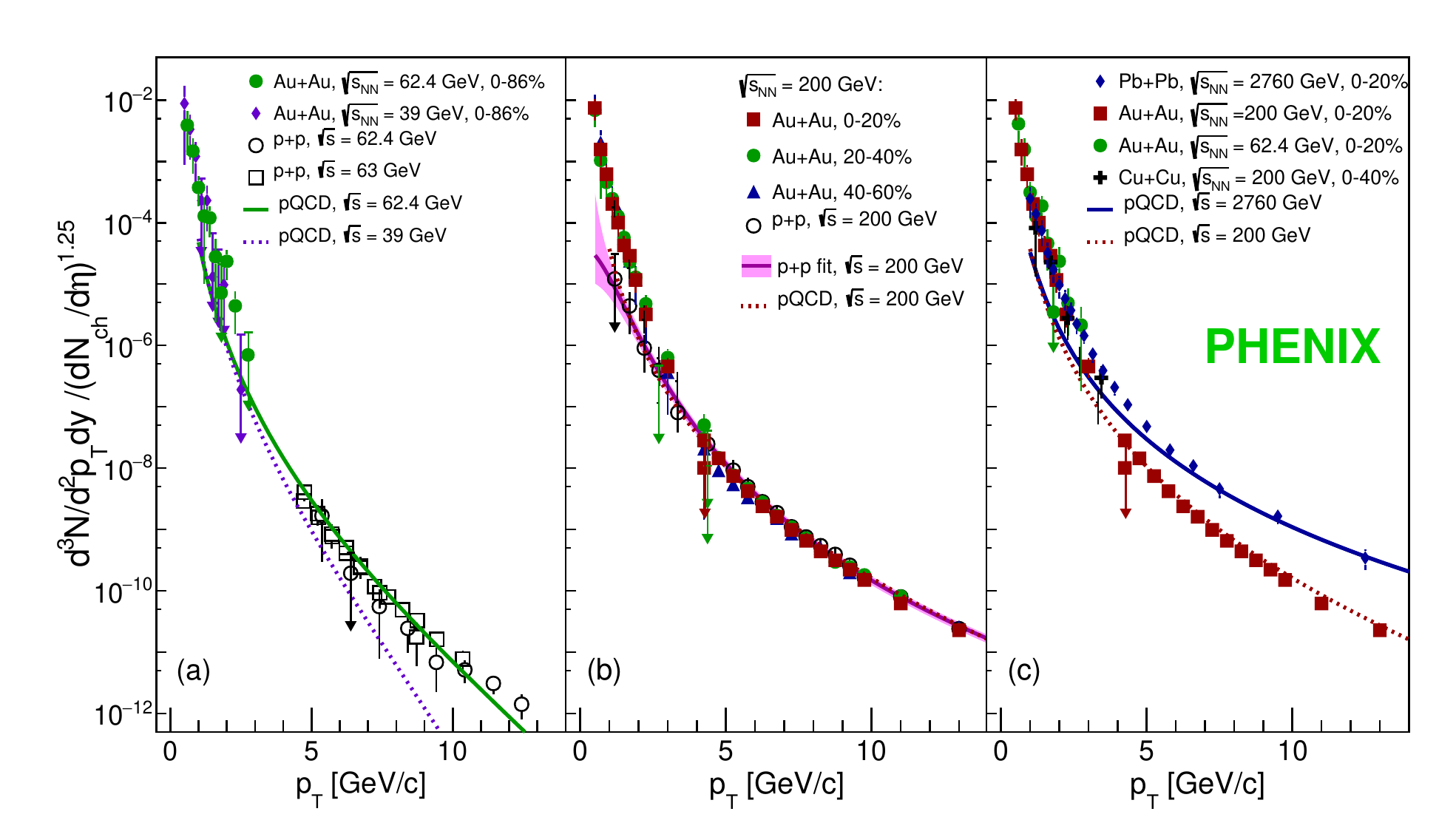}}
\end{center}
\vspace{-7.5mm}
\caption{This three-panel plot is from \cite{Adare:2018} showing the direct photon spectra normalized by $(dN_{ch}/d\eta)^{1.25}$. One can see the comparison for Au+Au data in minimum bias collisions at 62.4 GeV and 39 GeV (a); for Au+Au data in three centrality bins at 200 GeV (b); and for different A+A systems at four beam energies (c). Panels (a) and (b) also show p+p data, and all panels show perturbative QCD calculations at respective energies. All error bars are the quadratic sum of the systematic and statistical uncertainties. Uncertainties on $dN_{ch}/d\eta$ are not included. All normalized data, p+p fit, pQCD curves are from \cite{Adare:2018}, and they are obtained with Au+Au data from \cite{Bannier:2014,Adare:2008qk,Afanasiev:2012}, Pb+Pb data from \cite{Adam:2016}, p+p data at 200\,GeV from \cite{Adare:2013}, at 62.4\,GeV from \cite{Angelis:1980}, at 63\,GeV \cite{Angelis:1989,Akesson:1990}, the empirical fit to the p+p data at 200\,GeV from \cite{Adare:2018}, the pQCD calculations at different beam energies from \cite{Paquet:2015lta,Paquet:2015}, and the data on $dN_{ch}/d\eta$ from \cite{Adare:nch,Aamodt:2011nch}.
}
\label{fig:fig_scaling2}
\end{figure}

Thereby, we scale the direct photon yield by $(dN_{ch}/d\eta)^{\alpha}$, which for a given \sqn is equivalent to $N_{coll}$. For example, taking the photon spectra in minimum bias Au+Au collisions at 62.4 and 39\,GeV with pQCD curves from Fig.\,\ref{fig:fig_large}, and normalizing them by $(dN_{ch}/d\eta)^{\alpha}$, one can see that the data fall on top of each other at low-$p_{T}$ as shown in the panel (a) of Fig.\,\ref{fig:fig_scaling2}. At high-$p_{T}$ the p+p data coincide with the pQCD calculations
within the quoted uncertainties as expected. In the panel (b) all Au+Au data at 200\,GeV are on top of each other at high- and low-$p_{T}$, and at low-$p_{T}$ they are distinctly above the p+p data, fit and pQCD. In the panel (c) the data are compared for different \sqn from 62.4\,GeV to 2760\,GeV, scaled in the same way. And again all the data coincide at low-$p_{T}$, while at high-$p_{T}$ we see the expected difference with \sqn and $N_{coll}$ scaling. 
\vspace*{-0.4cm}

\begin{figure}[h!]
\vspace{-6.5mm}
\begin{center}
\hspace{0.0\textwidth}
\vspace{0.0\textwidth}
   {\includegraphics[width=0.475\textwidth]{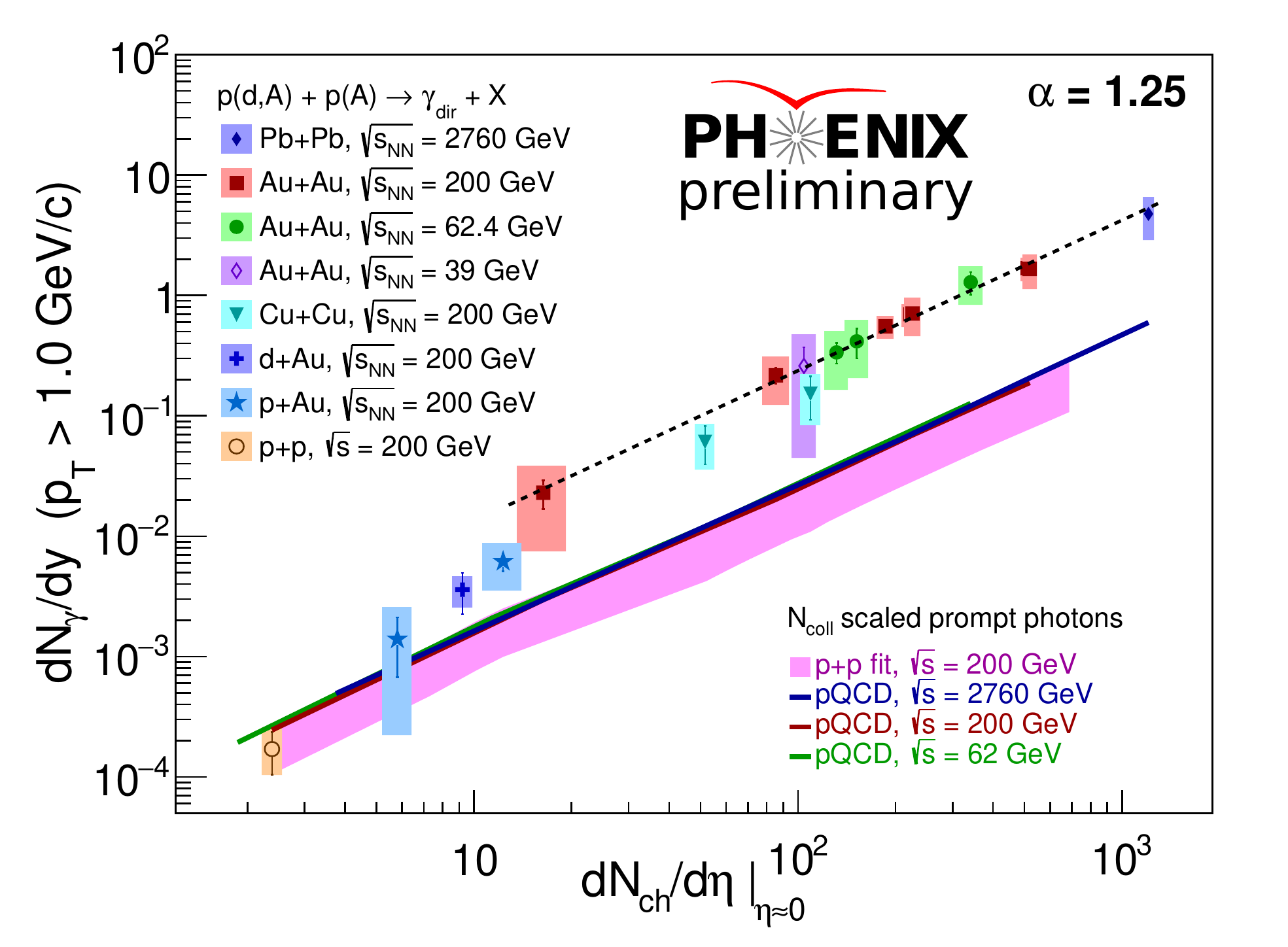} \label{fig:Fig5a}}
\hspace{0.00\textwidth}
\vspace{0.0\textwidth}
   {\includegraphics[width=0.475\textwidth]{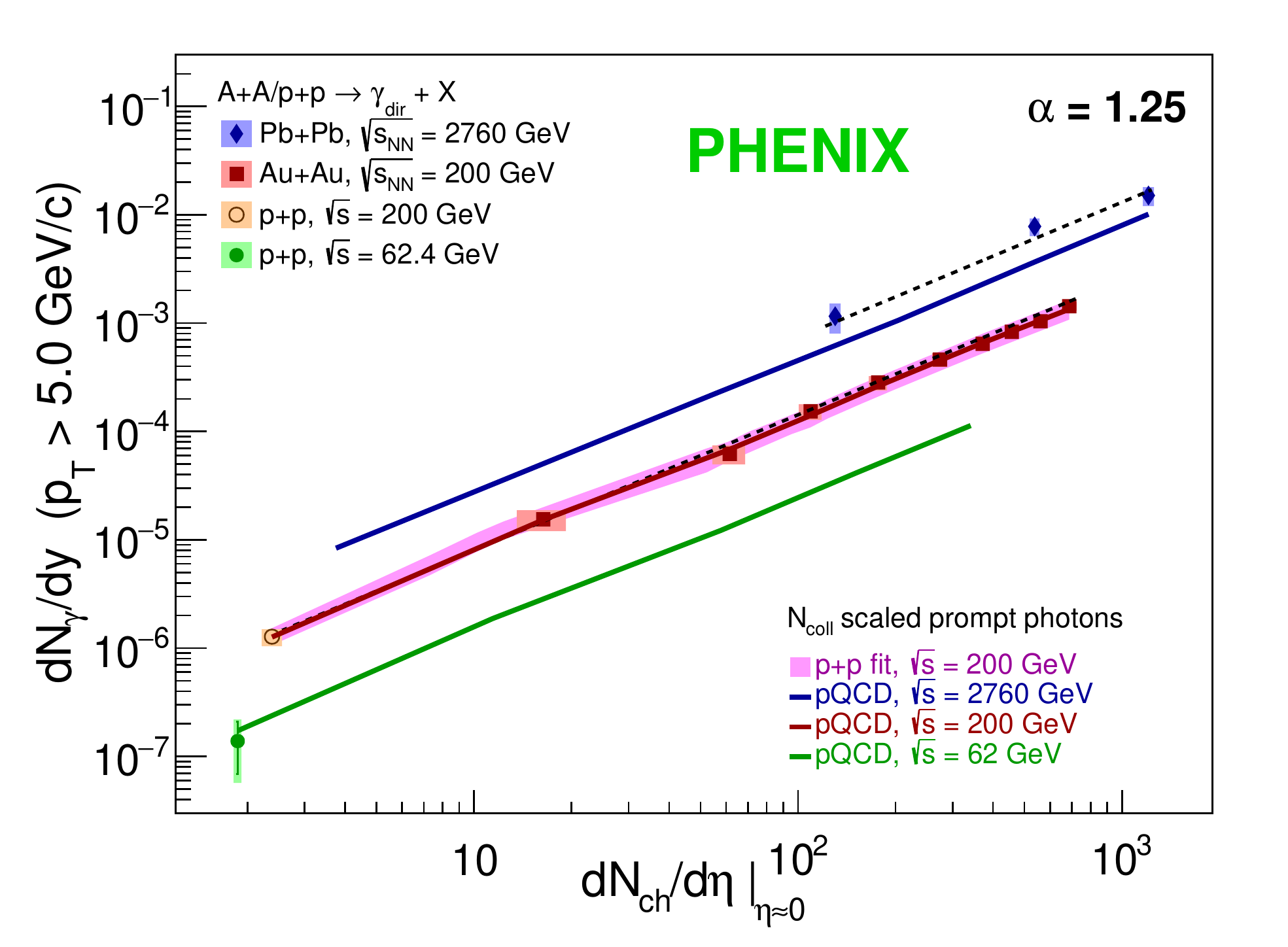} \label{fig:Fig5b}}
\end{center}
\vspace{-7.5mm}
\caption{
The left plot shows the data on the direct photon yield, integrated above 1.0\,GeV/c in $p_{T}$, vs. $dN_{ch}/d\eta$, for five A+A data sets at different collision energies and for one p+Au, one d+Au and one p+p data sets at 200\,GeV. Also, shown are the integrated yields of extrapolations (down to $p_{T} = 1$\,GeV/c) of the fit to p+p data and of the three different pQCD calculations scaled by $N_{coll}$. The right plot shows the yield integrated above 5.0 GeV/c in $p_{T}$, for two A+A data sets and two p+p data sets. The integrated yields from different pQCD calculations scaled by $N_{coll}$ are also shown. The black dashed lines are the power-law fits over the A+A data with a fixed $\alpha=1.25$ slope. In both plots the integration is carried out for the data, p+p fit and pQCD  curves from Fig.\,\ref{fig:fig_scaling2}.
}
\label{fig:fig_scaling3}
\end{figure}

\vskip -1.1truecm
In order to quantify the direct photon spectra, one can integrate the invariant yield over some $p_{T}$ threshold value. Then if we integrate above $p_{T} = 1$\,GeV/c, we obtain the left plot of Fig.\,\ref{fig:fig_scaling3}. This plot is another representation of the direct photon scaling, where the integrated yield of the large systems scales with $dN_{ch}/d\eta$ by the same power $\alpha = 1.25$, meaning that $dN_{\gamma}/dy$ grows faster than $dN_{ch}/d\eta$. It is interesting that the prompt photons described by the purple band and integrated pQCD curves have nearly the same slopes as that of the large systems. In the low multiplicity region one can see the gradually increasing trend of the integrated yield of the small systems, which seems to intersect with the trend from the large systems.

With the integration above high $p_{T} = 5$\,GeV/c, we get the right plot of Fig.\,\ref{fig:fig_scaling3}. Here the observed scaling behavior is expected \cite{Afanasiev:2012} (since $R_{AA} =1$), though we see that the slopes are the same as those in the left plot.

\vspace{-0.02\textwidth}
\section{Summary}
\label{sum}
The PHENIX collaboration has measured low momentum direct photons in Cu+Cu at \sqn = 200\,GeV, Au+ Au at \sqn = 62.4 and \sqn = 39\,GeV as well as in p+p and p+Au at \sqn = 200\,GeV. By compiling all the available results on small and large systems at various energies, PHENIX has found a surprising scaling behavior of direct photons in large systems, namely: at a given center-of-mass energy the low- and high-$p_{T}$ direct photon invariant yields from A+A collisions scale with $N_{coll}$; across different energies $N_{coll}$ is proportional to $dN_{ch}/d\eta$; meanwhile, for all energies the low-$p_{T}$ yield seems to scale like $(dN_{ch}/d\eta)^{\alpha}$. PHENIX has also discovered direct photon excess yield at low-$p_{T}$ in central p+Au collisions above $N_{coll}$ scaled p+p fit, which may originate from possibly existing QGP droplets in small central systems. Both trends described by the data seen in the left plot of  Fig.\,\ref{fig:fig_scaling3}, suggest the existence of a ``transition point"  between small and large systems.





\bibliographystyle{elsarticle-num}

\begin{thebibliography}{00}


\bibitem{Bannier:2016}
A.~Adare {\it et al.} (PHENIX Collaboration), Phys.Rev. C {\bf 94}, 064901 (2016) [arXiv:1509.07758 [nucl-ex]].

\bibitem{vanHees:2014ida}
H.~van Hees, M.~He and R.~Rapp, Nucl.\ Phys.\ A {\bf 933}, 256 (2015) [arXiv:1404.2846 [nucl-th]].

\bibitem{Paquet:2015lta} 
J.~F.~Paquet {\it et al.}, Phys.\ Rev.\ C {\bf 93}, no. 4, 044906 (2016)  [arXiv:1509.06738 [hep-ph]].

\bibitem{Kim:2016ylr} 
Y.~M.~Kim, C.~H.~Lee, D.~Teaney and I.~Zahed, Phys.\ Rev.\ C {\bf 96}, no. 1, 015201 (2017)  [arXiv:1610.06213 [nucl-th]].

\bibitem{Adare:cucu}
A.~Adare {\it et al.} (PHENIX Collaboration), [arXiv:1805.04066 [nucl-ex]].

\bibitem{Adare:2018}
A.~Adare {\it et al.} (PHENIX Collaboration), [arXiv:1805.04084 [nucl-ex]].

\bibitem{Adare:2013}
A.~Adare {\it et al.} (PHENIX Collaboration), Phys. Rev. C {\bf 87}, 054907 (2013) [arXiv:1208.1234 [nucl-ex]].

\bibitem{Adare:nch}
A.~Adare {\it et al.} (PHENIX Collaboration), Phys. Rev. C {\bf 93}, 024901 (2016) [arXiv:1509.06727 [nucl-ex]].

\bibitem{Aamodt:2011nch}
K.~Aamodt {\it et al.} (ALICE Collaboration), Phys. Rev. Lett. {\bf 106}, 032301 (2011) [arXiv:1012.1657 [nucl-ex]].

\bibitem{Bannier:2014}
A.~Adare {\it et al.} (PHENIX Collaboration), Phys.Rev. C {\bf 91}, 064904 (2015) [arXiv:1405.3940 [nucl-ex]].

\bibitem{Adare:2008qk}
A.~Adare {\it et al.} (PHENIX Collaboration), Phys. Rev. Lett. {\bf 104}, 132301 (2010) [arXiv:0804.4168 [nucl-ex]].

\bibitem{Afanasiev:2012}
S.~Afanasiev {\it et al.} (PHENIX Collaboration), Phys. Rev. Lett. {\bf 109}, 152302 (2012) [arXiv:1205.5759 [nucl-ex]].

\bibitem{Adam:2016}
J.~Adam {\it et al.} (ALICE Collaboration), Phys. Lett. B {\bf 754}, 235 (2016) [arXiv:1509.07324 [nucl-ex]];

\bibitem{Angelis:1980}
A.~L.~S.~Angelis {\it et al.} (CCOR Collaboration), Phys. Lett. B {\bf 94}, 106 (1980).

\bibitem{Angelis:1989}
A.~S.~Angelis {\it et al.} (CMOR Collaboration), Nucl. Phys. B {\bf 327}, 541 (1989).

\bibitem{Akesson:1990}
T.~Akesson {\it et al.} (AFS Collaboration), Sov. J. Nucl. Phys. {\bf 51}, 836 (1990).

\bibitem{Paquet:2015}
J.-F.~Paquet, Private communication, (2017).


\end{thebibliography}



\vspace{0.005\textwidth}

\end{document}